# Economics of Resilient Cloud Services

Brandon Wagner[1] and Arun Sood [1,2]
Department of Computer Science and International Cyber Center[1]
George Mason University, Fairfax, VA
SCIT Labs, Inc, Clifton, VA[2]

*Abstract*—Today's computer systems must meet and maintain service availability, performance, and security requirements. Each of these demands requires redundancy and some form of isolation. When service requirements are implemented separately, the system architecture cannot easily share common components of redundancy and isolation. We will present these service traits collectively as cyber resilience with a system called Self-Cleansing Intrusion Tolerance (SCIT). Further, we will demonstrate that SCIT provides an effective resilient cloud implementation making cost effective utilization of cloud's excess capacity and economies of scale. Lastly, we will introduce the notion of serverless applications utilizing AWS Lambda and how a stateless architecture can drastically reduce operational costs by utilizing cloud function services.

## I. INTRODUCTION

In our highly connected world, systems are expected to be online at all times while maintaining a high degree of security and performance. Developing systems that meet these standards require an extensive amount of time and money. The extra resources needed to develop high quality systems forces businesses to make trade-offs in performance, security, and cost. Resilience is a broad term that describes systems that meet these requirements. Haimes presents a comprehensive view of resilient systems [1]. We build on Haime's view and include resilient services in our definition. In our approach a resilient system and service is able to withstand a major disruption because of unknown events. The disruption's impact must be measured within acceptable degradation parameters and yield recovery to a known good state within an acceptable amount of time. This paper will outline strategies to achieve cyber resilience in the context of service availability, cyber-attacks, and reduced cost utilizing the Self Cleansing Intrusion Tolerance (SCIT) system.

### A. SCIT

The Self-Cleansing Intrusion Tolerance (SCIT) system works to increase a system's security by rotating servers periodically and restoring them to a clean image. SCIT improves security by decreasing the period of time in which a server is vulnerable to attack. Attackers usually remain in a compromised system for days, months, or even years. SCIT rotations can occur in minutes. The rapid frequency of rotations severely impacts an attacker's ability to persist in the system, exfiltrate data, and modify the system.

In order for an application to operate without issue on constantly rotating servers, the application must be sufficiently tolerant to availability failure. Usually this means offloading any state data to a separate application tier. For example, web-session data could be offloaded to a distributed caching tier such as Redis or a relational database. File upload and download services could be offloaded to a distributed data store such as AWS Simple Storage Service (S3), Google Cloud Storage, or a network file system (NFS) storage cluster.

In systems deploying Network Intrusion Detection Systems SCIT's rotation scheme is often an adequate substitute for a host-based intrusion detection systems (HIDS), thus reducing the performance overhead because of HIDS. However, the SCIT system does require an increased amount of redundancy to facilitate the rotations. There is a constant amount of time to provision a server with the clean image. During this constant time the system will need a running server online and a server running that is being provisioned. As the rotation period decreases, the amount of additional servers that need to be provisioned increases. Figure 1 shows a SCIT system with a rotation period of 1 hour. The next server to rotate into an active state must be provisioned before the previous server is shutdown.

This paper will focus on decreasing the cost required to operate the additional redundancy needed for a best practice, cloud-based security solution based on performing SCIT rotations. We will focus on using Amazon Web Services and Google Cloud Platform to compute the associated costs. Our baseline will take into account "on-demand" server resource costs in the two public clouds. We will then investigate the use of AWS's Spot Instances and GCP's Preemptive Instances to lower the operating costs.

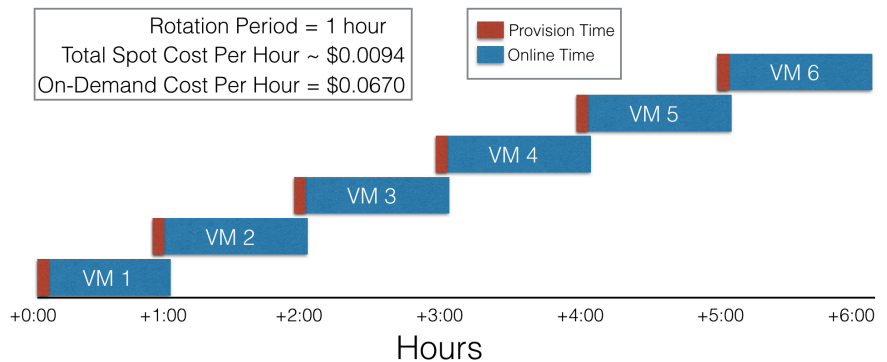

Figure 1: SCIT Virtual Machine Rotations

## II. STRATEGIES FOR RESILIENCE

### A. Redundancy

There are many vectors to consider when building resilient systems that usually include different forms of redundancy. We will describe redundancy of physical machines, virtual machines (VMs), geographic locations, and network/cloud infrastructures. Redundant components incur a monetary cost to utilize and maintain.

#### 1) Physical Servers, Virtual Machines, and Containers

In the simplest case, a resilient system will need multiple physical servers running the same software to provide a failover system in case of a transient fault in software or hardware failure. It may also be required to perform load balancing with redundant physical servers to prevent system overload that would affect availability and performance. Load balancing can mitigate a high volume of traffic, but it is still necessary to have a failover machine cluster in case of complete failure. Physical servers are often replaced with VMs to optimize costs, automate processes, and more quickly recover. VMs can greatly improve performance of a distributed and clustered system, but there must also be redundant physical machines to prevent a hardware failure from disrupting many VM instances. Containerization technologies such as Docker and rkt (pronounced "Rocket") should be treated similarly to VM instances in terms of requiring redundant physical hardware. There are still advantages to container technologies versus VMs in terms of portability, performance, and automation. Li et al. [2] expands on how to create high availability systems using a cloud provider. SCIT's architecture supports container ready applications that can benefit from the increased speed and portability with little cost. Containers can also be used to dramatically increase utilization, which can directly impact cost optimizations.

#### 2) Location Distribution

Location redundancy is a very important part of resilient systems in terms of performance and availability. Data center architectures usually contain some level of isolation and redundancy. Regions and Availability Zones are often used when discussing cloud architectures. A region is a group of availability zones within a relatively close proximity. AWS notes that availability zones are connected via private network lines where regions are connected via the Internet. Availability Zones can be thought of as a single data center or group of data centers where systems are isolated (dedicated power feed, dedicated network, etc.). The idea is that an availability zone can experience failure via a power or network outage and the other availability zones within the region can continue to operate because they do not rely on each other. There could be situations where increased load causes issues when multiple availability zones are unavailable which leads to a failure.

In order to build a resilient system, physical servers must be distributed across availability zones and regions. This reduces the reliance on a single data center or a group of data centers that could be affected by a natural disaster. Further, distributing a system to different regions through a content delivery network (CDN) can reduce the access time latency for users within that region.

#### 3) Provider Distribution

It is also important to consider the provider of your infrastructure. Even though both Amazon and Google operate massive infrastructures that would serve all of your system's needs, it is not ideal to go "all in" with a single cloud service provider. Cloud services are growing at an accelerated pace which leads to complex environments and a lot of software. Software tends to have bugs and can be vulnerable to attack or failure. Your system may be perfect, but if a cloud service provider is infiltrated or software services crash, your system's performance or availability could be adversely affected. Utilizing both AWS and Google Cloud Platform provides a stronger form of resilience that could be tolerant to uncontrollable cloud service failures. [3] suggests storing encryption keys on multiple cloud providers by using Shamir's secret sharing algorithm. There are software products and platform layers that aid system designers and implementers to more effectively utilize multiple clouds. Pivotal's Cloud Foundry platform-as-a-service (PaaS) can operate in multiple cloud environments to deploy systems. Netflix developed a continuous delivery (CD) tool called Spinnaker that aims to support multiple cloud environments including AWS, GCP, Microsoft's Azure, and Pivotal's Cloud Foundry PaaS.

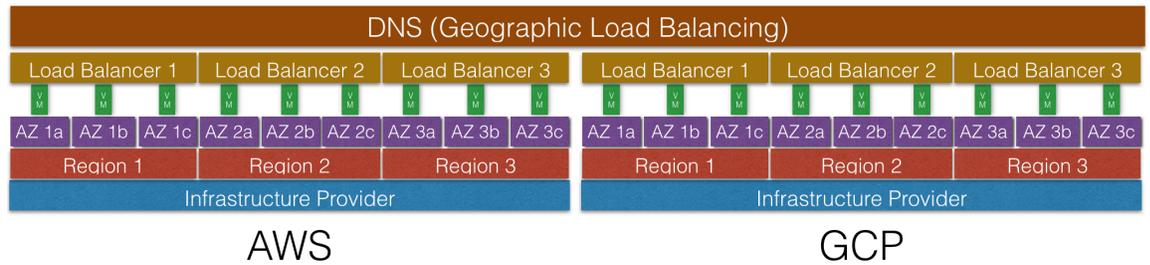

Figure 2: Multi-Provider, Availability Zone, and Regional Redundancy

### B. Diversity

Cyber attacks can cause crippling affects on a system's resilience. In order to defend a system from common vulnerabilities, an approach utilizing system diversity can help to segment attacks. Segmentation can work in tandem with redundancy to provide overall resilience. One diversity strategy that is simple to implement is server diversity. This type of diversity can be achieved by utilizing different infrastructure providers. GCP's servers are quite different from AWS's. More so, AWS's and GCP's managed services are implemented uniquely. For example, a vulnerability in an AWS load balancer may not be present in a Google Compute Engine Load Balancer. Diversity can be improved even more without a lot of complexity by utilizing different operating systems.

The SCIT rotation approach makes some of the low cost diversity approaches more effective. To illustrate this advantage, we will focus on two stronger forms of diversity to support resilient systems: Address Space Layout Randomization (ASLR) and N-Version Programming.

### 1) Address Space Layout Randomization (ASLR)

Address Space Layout Randomization (ASLR) is a defense mechanism at the operating system level to randomize the address space. This type of defense can mitigate return-oriented programming attacks introduced by buffer overflow exploits. ASLR has been integrated into most modern operating systems as a primitive defense. Shacham et al. [4], yields that ASLR is not particularly effective as a sole means for defense against buffer overflow exploits. However, they do note that ASLR can slow down propagation of worms. In the case of SCIT, any amount of additional work needed by the attacker can help to improve the security of the system as a whole. ASLR only needs to thwart the attack for the duration of a rotation period where SCIT will take over by taking the current server under attack out of service and instantiating a new server with a different memory mapping. Shacham et al., analyzed an implementation of a buffer overflow exploit to gain a remote shell on a system using ASLR. They were able to obtain the shell in about 200 seconds. A SCIT rotation period of 90 seconds would completely mitigate their attack.

Bittau et al. [5] notes that a return oriented programming attack (ROP) can be disrupted by applying ASLR to all executable segments and frequent re-randomization. Thus a small enough SCIT rotation period increases the resilience and security of the service. Further, gaining a remote shell is usually only the first step in an attack for purposes of recon and probing of systems deeper within the network. SCIT rotations would continually slow down the process of probing even if the initial exploit was executed successfully.

### 2) N-Version Programming

Another diversification strategy to enhance the resilience of a system is N-Version Programming. A majority of the literature on N-Version Programming recommends multiple isolated branches of the same business functionality. This is not practical to a commercial company from a cost perspective and maintainability. The SCIT rotation approach makes a weaker form of N-Version programming more effective. We recommend a weaker form of N-Version Programming suggested by Murphy et al. [6], where software is compiled with multiple compilers and varying compilation flags. The approach can be generalized to varying runtimes as well. This approach is easy to place within an existing development workflow with continuous integration (CI), continuous delivery (CD), and automated build systems.

Another form of weak N-Version Programming to facilitate diversity is varying containerized runtimes. You can use Docker's Dockerfile to define a different virtualized operating system to run on top of a host OS. This could involve using different libraries, compilers, and runtimes (including JRE versions). Docker containers are advantageous because of their portability across Linux machines. You can implement two levels of diversification by varying base OS distributions of the VM/physical server illustrated in figure 3. Diversification with N-Version programming can aid in slowing down vulnerability exploitation and sequential knowledge building for more extensive system attacks.

With this approach, each successive iteration of a SCIT VM will have the same functionality but a different footprint. This makes it more difficult for the attacker to exploit system vulnerabilities.

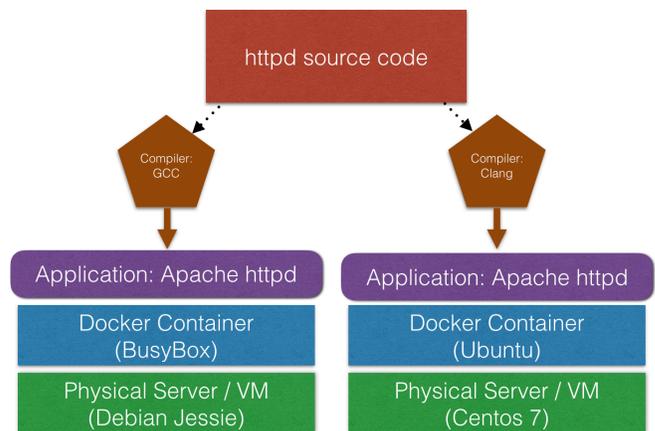

**Figure 3: Example of Diversifying Software**

## III. CLOUD PLATFORM PRICING MODELS

Cloud providers that offer servers and services on-demand have to predict usage to optimize their physical resources. This is a difficult task since on-demand usage lends itself to experimentation and the customer's own cost savings approaches by shutting down servers that are not being used. For example, many large-scale websites will run a base load to accommodate normal traffic and then provision additional servers when traffic increases. AWS and GCP both have to maintain a physical infrastructure to accommodate their customers' maximum scale-up. Most of the time, these extra resources are idle which led to a new pricing model for excess resources. AWS calls these resources Spot Instances and GCP calls them Preemptive Instances.

### A. AWS Spot Market

AWS's spot instances operate within a supply and demand market where customers programmatically bid on resources at a discounted rate. If demand increases within the market or if supply decreases, the prices will go up. If the price rises above a customer's maximum bid, the server will be shutdown after a two-minute warning. Ben-Yehuda et al., suggests that AWS Spot pricing is not actually determined by bidding. They concluded that Amazon sets a hidden reserve price based on their supply of resources and prediction of utilization [7]. There are conflicting reports that conclude bidding prices are input into the pricing function. Either conclusion does not affect our strategy in optimizing an overall system's cost.

### B. GCP Preemptive Instances

GCP's preemptive instances operate at a higher level of abstraction than AWS's spot instances. GCP hides the market change from its customers, which makes the process simpler. GCP notes that instances can be utilized at up to a 70% discount. GCP only sends a 30 second warning when an instance is preempted, or terminated, compared to AWS's two minute warning.

### C. AWS and GCP Billing Periods

Another notable difference between AWS and GCP are their pricing periods. AWS bills per hour and GCP bills per minute. For normal applications, the billing period does not make a significant difference, but SCIT's rotation cycle causes this issue to be significant. For example, if a customer boots up an AWS VM for 10 minutes, then they are charged for the full hour. If a GCP VM were booted for 10 minutes, the customer would only be charged for the 10 minutes it was online. GCP has a 10 minute minimum billing period. AWS combines instance duration per hour so that you are not billed per individual instance within an instance type. For example, a customer can start two identical EC2 instances, run them for 10 minutes, and they are only charged for 1 hour considering the 20 minutes of usage. Spot instances follow a different model since you are bidding for a specific instance. Therefore, if a customer boots up two spot instances, each for 10 minutes, they are billed 1 hour for each spot instance, totaling two hours of use. A spot instance is charged for 1 hour after booting, but if a spot instance is stopped before the two-hour mark, the customer is only charged for 1 hour of use [8]. Together with SCIT concepts we will use this strategy to optimize servers and system rotation periods when using AWS spot instances.

### D. AWS Reserved Instances

AWS offers reserved instances to customers that agree to use a VM for a certain time period in exchange for a discount per hour. Right now, the time frames are one year or three years. The customer can elect to pay all upfront, partial upfront, or no upfront (only for the one year reserved). An all-upfront, three year reserved instance offers the largest discount while the one-year no upfront offers the smallest discount.

### E. GCP Sustained Discount

GCP does not have an equivalent offering like AWS reserved instances. However, GCP offers a discount for sustained usage per month. Sustained usage is calculated by breaking a month up into four parts. Discounted rates are applied per sector. The discounts vary per instance, but a 20%-30% discount can be achieved with sustained usage. GCP calculates all VMs in use (inferred instances), which allows customers to start and stop VMs while still benefiting from the sustained usage discounts [9].

## IV. ECONOMIC OPTIMIZATION WITH SCIT

### A. SCIT Process Optimizations

SCIT lowers the security overhead by reducing the amount of time necessary to perform security functions. In a conventional system (no-SCIT), a lot of the time is needed to investigate and respond to a security event because the system is still operating in a production environment. After an attack, a system usually has to failover or rollback to a safe state. In SCIT the VM is regularly decommissioned and thus easier to analyze.

### B. SCIT Operational Optimizations with AWS

A SCIT system can be effectively operated while drastically lowering the cost of redundancy in the AWS cloud environment. Our sample operational scenario is a Spring Boot application that has file upload and download functionality to AWS S3.

The application will operate on AWS with all spot instances. Since spot instances can be reclaimed by AWS at anytime and are not guaranteed to be available, we will use a pool of four on-demand hot swappable instances. Each instance takes three minutes to provision with the Spring Boot application. Our on-demand instances will be provisioned and then stopped. On-demand instances that are stopped do not incur service charges except for elastic block storage (EBS) volumes. For this application, we only need an EBS volume large enough to hold our Spring Boot application, so the charge will be insignificant.

| Type | 1 hour | 30 days | 365 days |
|---|---|---|---|
| Spot (2) *Base Load | $0.0600 | $43.200 | $525.60 |
| Spot (5) *Peak Load | $0.15 | $18.00 | $219.00 |
| Total | | $61.20 | $744.60 |

Figure 4: AWS Spot Instance Cost (rotating ~ 1/hr)

| Type | 1 hour | 30 days | 365 days |
|---|---|---|---|
| On-Demand (2) *Base Load | $0.203 | $146.16 | $1,778.28 |
| On-Demand (5) *Peak Load | $0.51 | $60.90 | $740.95 |
| Total | | $207.06 | $2519.23 |

Figure 5: AWS On-Demand Instance Cost (rotating ~ 1/hr)

Our sample file upload and download application is deployed within a SCIT-like system rotating approximately every one hour and runs on M4.Large AWS EC2 instances. We accounted for a peak load period of 4 hours per day (2 hours in the morning and 2 hours in the evening), where spot instances will be booted to accommodate the extra volume. The rest of the time, we will keep two spot instances available to serve requests. This deployment scenario will cost approximately $61.20 per month assuming an average spot price of $0.03 and peak load requiring an additional 5 spot instances. Figure 4 shows the breakdown. A comparable deployment scenario utilizing only on-demand instances would cost around $259.20 per month. The redundancy while provisioning on-demand instances incurs a $14.40 charge per month for the two extra hours needed. Note that there is no additional cost for the redundant spot VMs because we are utilizing AWS's second hour grace period for spot instances. Figure 5 shows the breakdown of cost for an on-demand only deployment. Using only on-demand instances is four times more expensive than using spot instances. If the system is already compatible with SCIT (meaning fault tolerant to reboots and statelessness), then we can achieve these cost savings with little effort. As the system grows, the cost savings will also grow since we are able to scale up with spot instances during peak load.

AWS's billing period causes issues when trying to achieve a system rotation period less than 1 hour with spot instances. It is more cost efficient to use reserved and on-demand instances when trying to achieve very short rotation periods. AWS on-demand instances with a five-minute rotation period will be billed two times the amount of a single instance per hour. This is because of the redundancy needed to provision an instance before it is online. We are able to limit the costs with the one-hour rotation period by only leaving the instance online for 55 minutes, which allows five minutes for provisioning.

### C. SCIT Operational Optimizations with GCP

GCP's prices per instance hour are very similar to AWS's. Google's N1-Standard-2 is comparable to AWS's M4.Large in terms of specifications. The preemptible pricing is $0.03 per hour, which is the same as AWS's average spot price for the M4.Large. The N1-Standard-2 is priced at $0.10 per hour without sustained usage and $0.07 with full-sustained usage. AWS's pricing for the M4.Large is $0.12, so Google's pricing is slightly cheaper.

GCP's billing period is per minute with a 10-minute minimum after the first two minutes. This billing model allows us to optimize our costs when we set our rotation period to less than one hour. We can achieve a similar cost as the AWS usage scenario depicted in Figure 4 with a 10-minute rotation period. Assuming a three-minute provision time, we can expect to accumulate an additional 27 minutes per hour or 648 minutes per day because of the redundancy needed to rotate and provision VMs. For preemptible instances the cost to operate 648 minutes is an additional $0.324 per day per instance. On-demand instances will incur an additional $0.756 per day.

| Type | 1 hour | 30 days | 365 days |
|---|---|---|---|
| Preemptible (2) *Base Load | $0.087 | $62.64 | $762.12 |
| Preemptible (5) *Peak Load | $0.22 | $26.10 | $317.55 |
| Total | | $88.74 | $1079.67 |

Figure 6: GCP Preemptible Instance Cost (rotating ~ 10 mins)

| Type | 1 hour | 30 days | 365 days |
|---|---|---|---|
| On-Demand (2) *Base Load | $0.2400 | $187.200 | $2,277.60 |
| On-Demand (5) *Peak Load | $0.60 | $72.00 | $876.00 |
| Total | | $259.20 | $3153.60 |

Figure 7: GCP On-demand Instance Cost (rotating ~ 10 mins)

## V. SERVERLESS EXECUTION

SCIT focuses on the security of servers since this is a major security concern to organizations. Cloud providers have given customers the tools to think of servers as immutable infrastructure. We can just as easily provision a new VM with an image modification than to change something on the existing machine. Cloud providers are now pushing the idea further with managed code execution. AWS offers a service called Lambda that will execute code within its compatible runtimes, which includes Java, NodeJS, and Python. The customer simply uploads an archive of code and Lambda will take care of provisioning a virtual container to run the code. The code can read and write files when executing, but the environment will be reset on every execution and will not be running at all when it is not called. The notion of a serverless operating environment is very attractive in terms of security, cost, and architecture.

First, there is a much smaller attack surface when executing on a platform that does not allow you to open ports, run multiple applications, and that is not online all of the time. The environment is similar to what SCIT achieves in server operation by resetting the virtual space. However, AWS Lambda and other similar services do not reboot the actual server running the containers as SCIT would. This means that if an attacker was able to break out of the container and assume control of the host machine, then all containers running code on the machine would be vulnerable to attack. Second, the billing model for the Lambda service is much different than EC2. Lambda bills per 100 milliseconds of execution time [10]. The Lambda function is triggered by some external source which could be an S3 file upload, an HTTP request to AWS's managed API Gateway, or a time clock for cron-like executions. The only resource that can be customized when running on Lambda is RAM which ranges from 128MB to 1.5GB which also affects the CPU speed AWS provisions for you. AWS has made the first 3.2 million compute seconds free per month for 128MB Lambdas. For a lot of applications, architecting for Lambda compatibility would significantly reduce the cost required to operate, even when comparing against spot or preemptible pricing. After the allotted free Lambda executions, pricing begins at $0.000000208 per 100 milliseconds. Figure 8 shows the estimated cost for operating a similar application to figures 4-7. It uses a 512 MB Lambda function utilized 100% of the time with 1 request/second. AWS does not charge for the first 800,000 execution seconds for a 512 MB function and the first 1,000,000 function requests, which makes Lambda about 75% cheaper than AWS spot instances. Lastly, the architecture that Lambda promotes is similar to the fault tolerant engineering that SCIT requires which means applications can be easily transferred between SCIT and Lambda without a lot of architectural overhead. The application must be stateless and offload service tiers for storage and, in Lambda's case, HTTP endpoints. The code is forced into being loosely coupled since each application is broken up into separate Lambda functions that cannot communicate except through external application tiers.

AWS is in charge of managing its fleet of servers to operate the virtual environments that Lambda executes from. In order to operate this fleet effectively in terms of cost and performance, AWS Lambda functions are limited to a five-minute execution time. Patching, code isolation, and immediate execution with no provisioning time are all provided by AWS.

GCP launched a service called Google Cloud Functions that aims to compete with AWS Lambda. The service is currently in an Alpha stage, as of June 2016, that requires filling out a form and acceptance by Google to trial the service [11].

## VI. CONCLUSION

We cannot avoid failures or successful attacks all of the time, thus our push is to design resilient and secure systems in tandem. This presents a big challenge in the dynamic cloud ecosystem. SCIT's requirements for redundancy and architectural constraints of statelessness have an impact on how a system is engineered. These design decisions lead to a more resilient system by engineering around the idea that failure will occur often. SCIT's approach is general to servers and can be implemented on a wide variety of applications. AWS and GCP's excess resource capacity can be used to significantly reduce the cost of operating a system that can withstand failure. We were able to provide redundancy while reducing the cost of operating an application without redundancy. Further, managed code execution services such as AWS Lambda and GCP's Google Cloud Functions can significantly reduce the cost of operating a resilient system even in comparison to spot and preemptible VMs. System design is usually composed of tradeoffs between security and performance because of the cost of redundancy and complexity of managing redundant systems. This paper shows that creating a resilient system in terms of security, availability, and performance can yield significant economical advantages to organizations.

| Type | 1 hour | 30 days | 365 days |
|---|---|---|---|
| Execution Time Cost *Base Load | $0.00 | $14.95 | $182.95 |
| Requests Cost (assuming 1 req/sec) | $0.00 | $0.32 | $3.91 |
| Total | | $15.26 | $186.85 |

Figure 8: AWS Lambda Cost

In this paper we have discussed examples of applying SCIT to two public clouds. However, SCIT and our analysis can be readily extended to hybrid systems like private and public clouds, clouds from multiple vendors, and a mix of on-premise systems and cloud environments. This paper demonstrates cost effective solutions to meet customer resiliency requirements utilizing the SCIT approach.

## Acknowledgements

We thank the anonymous reviewers. Their reviews improved this paper. This research was partially supported by the Office of Naval Research contract N000141310017.